\documentclass[a4paper,12pt]{article}

\usepackage{setspace}
\usepackage{amsmath,amssymb,amsfonts}
\usepackage{graphicx}
\usepackage{color}
\makeatletter
\@addtoreset{equation}{section}
\renewcommand{\theequation}{\thesection.\@arabic\c@equation}
\makeatother
\usepackage{hyperref}
\usepackage{cite}
\usepackage{caption}
\usepackage{ulem}

\definecolor{red}{rgb}{1,0,0}
\definecolor{green}{rgb}{0,1,0}
\definecolor{blue}{rgb}{0,0,1}
\definecolor{darkblue}{rgb}{0,0,0.5}
\definecolor{lightblue}{rgb}{.5,.5,1}
\definecolor{lightgray}{gray}{.87}
\definecolor{Dark}{gray}{.20}
\definecolor{pink}{rgb}{.95,0.82,0.92}
\definecolor{yellow}{rgb}{1,1,0}
\definecolor{lightyellow}{rgb}{1,1,.5}
\definecolor{purple}{rgb}{0.7,0,0.85}
\definecolor{darkgreen}{rgb}{0,0.5,0}
\definecolor{orange}{rgb}{0.8,0.2,0.2}
\def \be {\begin{equation}}
\def \ee {\end{equation}}
\def \bea {\begin{align}}
\def \eea {\end{align}}
\def \nn {\nonumber}

\def \rr {\raise.35ex\hbox{\small $\prime$}\kern-.17em{\mbox{\large $\imath$}}}
\def \del {\partial}
\def \dels {\partial\kern-.5em / \kern.5em}
\def \As {{A\kern-.5em / \kern.5em}}
\def \Ds {D\kern-.7em / \kern.5em}

\def \Lam {\Lambda}

\def \th {\theta}

\newcommand{\detail}[1]{}

\newcommand{\hide}[1]{}

\newcommand{\remove}[1]{}
\newcommand{\change}[1]{#1}
\newcommand{\explanation}[1]{}

\newcommand{\new}[1]{{\color{blue}#1}}

\begin{document}

\pagestyle{plain}


\begin{titlepage}
\vspace*{-10mm}   
\baselineskip 10pt   
\begin{flushright}   
\begin{tabular}{r}
\end{tabular}   
\end{flushright}   
\baselineskip 24pt   
\vglue 10mm

\begin{center}

\noindent
\textbf{\LARGE
Distance between collapsing matter and apparent horizon in evaporating black holes
}
\vskip20mm
\baselineskip 20pt

\renewcommand{\thefootnote}{\fnsymbol{footnote}}
{\large
Pei-Ming Ho${}^{a}$%
\footnote[2]{e-mail: pmho@phys.ntu.edu.tw},
Yoshinori Matsuo${}^{b}$%
\footnote[3]{e-mail: matsuo@het.phys.sci.osaka-u.ac.jp}
and 
Yuki~Yokokura$^c$
\footnote[4]{e-mail: yuki.yokokura@riken.jp}
}
\renewcommand{\thefootnote}{\arabic{footnote}}

\renewcommand{\thefootnote}{\arabic{footnote}}

\vskip5mm

{\it  
${}^{a}$
Department of Physics and Center for Theoretical Physics, \\
National Taiwan University, Taipei 106, Taiwan,
R.O.C. 
\\
\vskip 3mm
${}^{b}$
Department of Physics, Osaka University, \\
Toyonaka, Osaka 560-0043,
Japan
\\
\vskip 3mm
${}^{c}$
iTHEMS Program, RIKEN, Wako, Saitama 351-0198, Japan
}

\vskip 25mm
\begin{abstract}

Assuming that the vacuum energy-momentum tensor
is not exceptionally large,
we consider 4D evaporating black holes with spherical symmetry
and evaluate the proper distance $\Delta L$ between the time-like apparent horizon
and the surface of the collapsing matter
after it has entered the apparent horizon.
We show that $\Delta L$ can never be larger than
$\mathcal{O}(n^{3/2}\ell_p)$
when the black hole is evaporated to $1/n$ of its initial mass,
as long as $n \ll a^{2/3}/\ell_p^{2/3}$
(where $a$ is the Schwarzschild radius and $\ell_p$ is the Planck length). 
For example,
the distance between the matter and the apparent horizon
must be Planckian at the Page time.

\end{abstract}
\end{center}

\end{titlepage}

\pagestyle{plain}

\baselineskip 18pt

\setcounter{page}{1}
\setcounter{footnote}{0}
\setcounter{section}{0}


\newpage

\section{Introduction}
It has been noted \cite{Mathur:2009hf} that,
if the information paradox of black holes is resolved by
converting all information of the collapsing matter 
into the Hawking radiation,
there must be high-energy events around the horizon
(such as the firewall \cite{firewall,firewall-B}).
However, if the collapsing matter is already far inside the horizon,
even a firewall around the horizon is still not enough
unless there are nonlocal interactions at work.

The question we want to answer in this paper is the following:
How far is the collapsing matter under the apparent horizon
when the black hole evaporates to a certain fraction of its initial mass,
say, one half at the Page time?

For 4D spherically symmetric evaporating black holes,
we consider a generic class of vacuum energy-momentum tensor 
without exceptionally large components.
(The conventional model of black holes
\cite{Davies:1976ei,Christensen:1977jc,Brout:1995rd}
is included.)
The back-reaction of the vacuum energy-momentum tensor
to the near-horizon geometry is taken into consideration
in the semi-classical Einstein equation.
The general solution for the near-horizon geometry is consistent with
related works on the conventional model
(see e.g. Ref.\cite{Massar}).
The light-cone coordinates $(u, v)$ is used
so that the causal structure is more manifest.

We prove that,
due to a robust exponential form of the redshift factor inside the apparent horizon,
the proper distance $\Delta L$ between the apparent horizon and
the collapsing matter
(after it has fallen inside the apparent horizon)
is never larger than $n^{3/2}\ell_p$
when the black hole is $1/n$ of its initial mass.
This estimate is valid until the black-hole mass is an extremely small fraction
$1/n \sim \mathcal{O}(\ell_p^{2/3}/a^{2/3})$ of its initial mass.
\footnote{For the vacuum energy-momentum tensor defined by a 2D massless scalar field \cite{Davies:1976ei},
the Planck-scale proper distance between apparent horizon and collapsing matter
was already argued in Ref.\cite{HoMatsuoYokokura}.
} 
This conclusion reveals an important feature about
the geometry under the apparent horizon.
In line with Refs.\cite{Baccetti:2018otf,Baccetti:2018qrp},
our result questions the validity of an effective theory with a cutoff scale $\Lam$
lower than the Planck mass $M_p$.

\section{Assumptions}\label{assumptions}

We start by listing all of our assumptions.

\begin{figure}[h]
\vskip-2em
\center
\includegraphics[scale=0.5,bb=0 0 300 450]{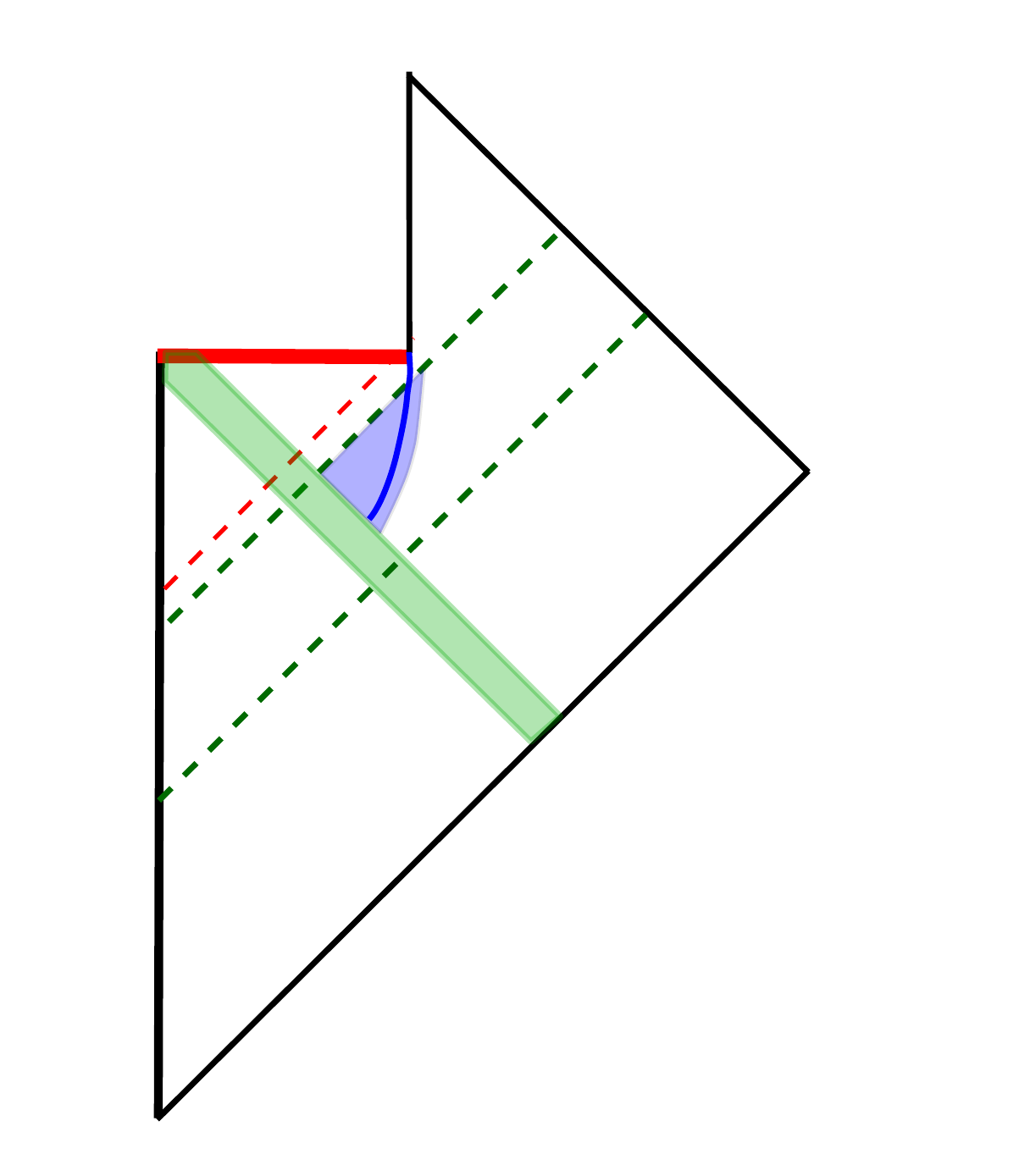}
\includegraphics[scale=0.45,bb=0 0 320 450]{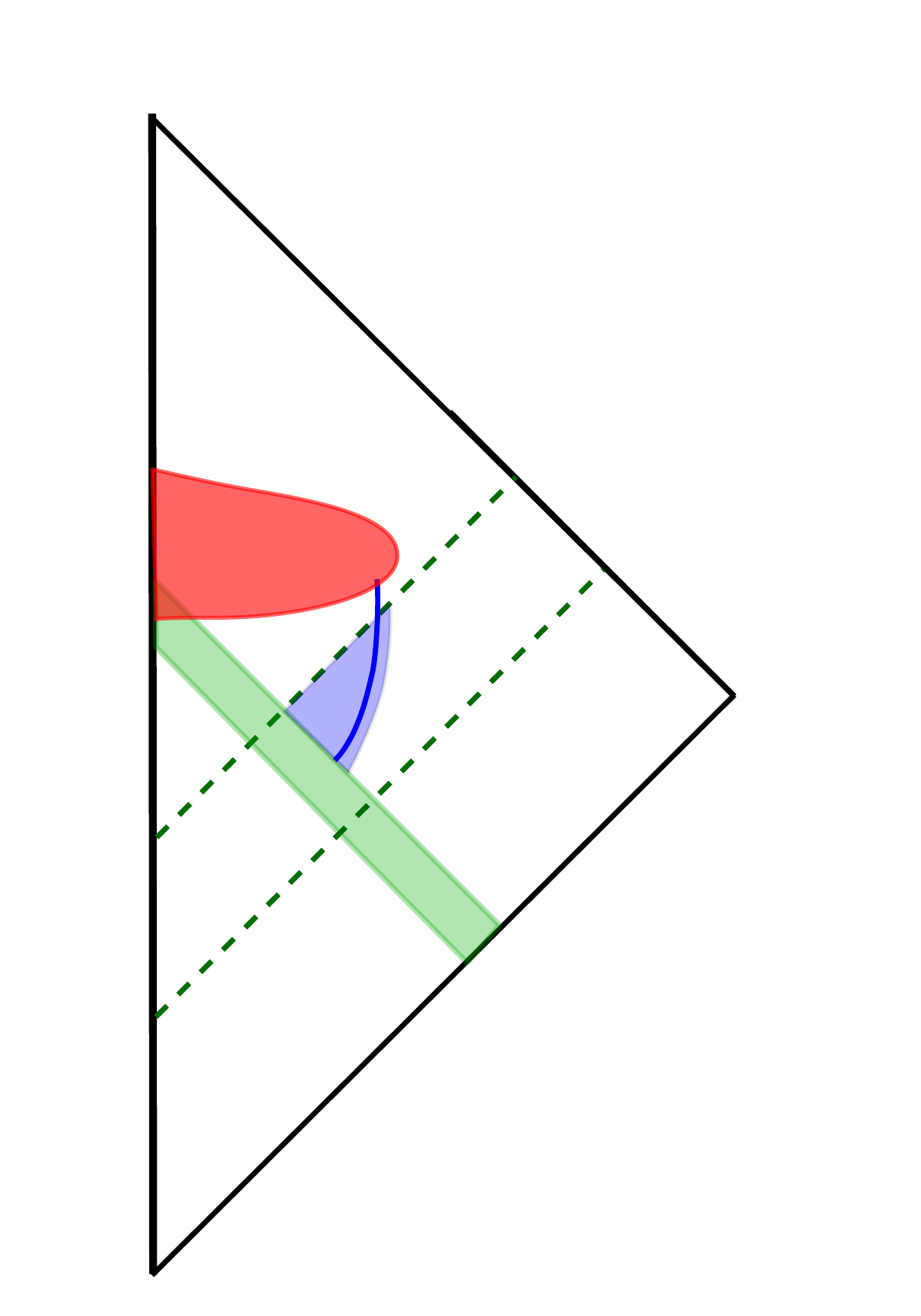}
\includegraphics[scale=0.57,bb=10 30 300 450]{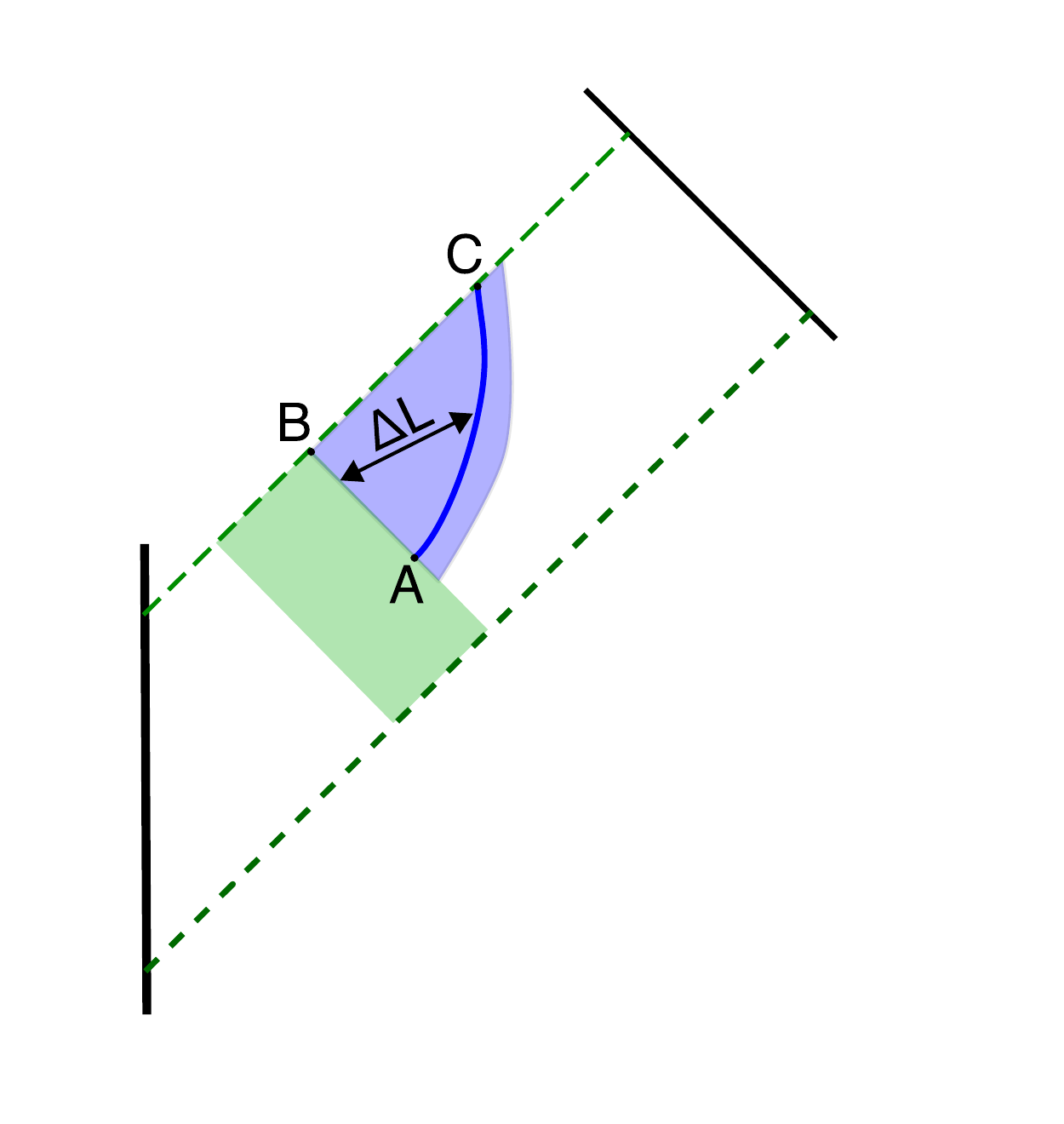}
\vskip0em
\hskip0cm (a) \hskip4.5cm (b) \hskip4.5cm (c)
\vskip1em
\caption{\small 
(a) Penrose diagram
for the conventional model.
The apparent horizon (blue curve) is time-like
outside the collapsing matter (thick green strip)
at the speed of light.
We focus on the spacetime (between the green dash lines)
not long before the apparent horizon emerges,
and before the black hole is completely evaporated,
from the viewpoint of a distant observer.
The event horizon (red dash line) is irrelevant.
(b)
The singularity at $r = 0$ is replaced by
a Planckian region (red blob) in the UV theory.
The region between the green dash lines remains the same.
(c)
The region of interest (between the green dash lines) excerpted from (a) and (b).
The ``near-horizon region'' defined in Sec.\ref{NHR} is the (blue shaded) region
bounded by the following 3 curves:
(1) the outer surface of the collapsing matter (thick green strip) from A to B,
(2) the outgoing null curve (dashed green line) from B to C,
(3) a curve outside but close to the time-like apparent horizon in vacuum from C to A.
We will study the proper length $\Delta L$ of a monotonic radial curve
connecting an arbitrary point on the surface of the collapsing matter between A and B
and another arbitrary point on the apparent horizon between A and C.
An example of such a curve (double arrowhead) is shown.
}
\label{Penrose}
\vskip1em
\end{figure}

\begin{enumerate}

\item
\label{macroscopic-black-hole}
{\bf Macroscopic evaporating black hole} \\
The Schwarzschild radius $a(t) \equiv 2G_N M(t)$
is much larger than the Planck length $\ell_p \equiv \sqrt{\hbar G_N}$:
\be
a(t) \gg \ell_p.
\ee
The time-evolution equation for $a(t)$
\be
\left|\dot{a}\right| \sim \mathcal{O}(\ell_p^2/a^2),
\label{dota}
\ee
for the time $t$ of a fiducial observer far from the black hole, 
so that the time for the black hole to evaporate to
a fraction ($<1/2$) of its initial mass is $\mathcal{O}(a^3/\ell_p^2)$.

\item
\label{spherical-symmetry}
{\bf Spherical symmetry} \\
The most general spherically symmetric 4D metric is
\begin{align}
ds^2 = - C(u, v) du dv + r^2(u, v) d\Omega^2,
\label{metric}
\end{align}
where $u$ and $v$ are the outgoing and ingoing light-cone coordinates,
$r(u, v)$ is the areal radius
and $d\Omega^2$ is the metric of a unit 2-sphere.

\item
\label{einstein-equation}
{\bf Semi-classical Einstein equation} \\
The semi-classical Einstein equation
\begin{align}
G_{\mu\nu} = \kappa \langle T_{\mu\nu} \rangle
\label{SCEE}
\end{align}
holds since we will focus only on spacetime regions
where the curvature is small.
Here,
$\kappa \equiv 8\pi G_N$
and $\langle T_{\mu\nu} \rangle$ is the expectation value
of the quantum energy-momentum tensor
for a given quantum state.
Since $\langle T_{\mu\nu} \rangle$ in vacuum
is proportional to the Planck constant $\hbar$,
the quantum correction to the Einstein equation
is proportional to $\mathcal{O}(\ell_p^2)$.

\item
\label{Schwarzschild-approx}
{\bf Schwarzschild approximation} \\
The near-horizon geometry is expected to deviate from
the classical solution due to quantum corrections,
but it should be smoothly connected to the classical metric at large distances. 
Within a time $\Delta t \sim \mathcal{O}(a)$ (see eq.\eqref{dota})
in a neighborhood where 
\be
r - a \gg \mathcal{O}\left(\frac{\ell_p^2}{a}\right),
\label{r-a}
\ee
the spacetime geometry is well approximated
by a Schwarzschild solution:
\begin{align}
C = 1 - \frac{a}{r}, 
\qquad
- \del_u r = \del_v r = \frac{1}{2}\left(1-\frac{a}{r}\right).
\label{Sch-sol}
\end{align}
Here we have used the Eddington light-cone coordinates $(u, v)$.

\item
\label{EMT-bounds}
{\bf Bounds on energy-momentum tensor} \\
The vacuum energy-momentum tensor is assumed to satisfy
\begin{align}
\langle T^{\mu}{}_{\mu}\rangle \lesssim \mathcal{O}\left(\frac{1}{\kappa a^2}\right),
\qquad
\langle T^{\th}{}_{\th}\rangle \lesssim \mathcal{O}\left(\frac{1}{\kappa a^2}\right)
\label{pressure-bound}
\end{align}
in the near-horizon region.
These bounds are much weaker than the condition of uneventful horizon
for the conventional model
\cite{Davies:1976ei,Christensen:1977jc,Brout:1995rd}.
This quantum effect is important for the apparent horizon.

\item
{\bf Apparent horizon}
\label{TH} \\
We assume that the apparent horizon exists.
\footnote{
The appearance of the apparent horizon has been viewed
as a definition of black holes
\cite{Booth:2005qc,Thornburg:2006zb,Faraoni:2013aba}.
Note that we are now considering a different situation from Refs.\cite{Kawai:2013mda,KMY}.
}
(See Fig.\ref{Penrose} and note its difference from the event horizon,
which is irrelevant to our discussion below.)
Due to the ingoing negative energy of the vacuum fluctuation, 
the outer trapping horizon is time-like as it emerges outside the collapsing matter
\cite{Frolov:1981mz,Roman:1983zza,Hayward:2005gi},\cite{Baccetti:2018otf,Baccetti:2018qrp},\cite{Ho:2019kte}.
\end{enumerate}

The quantum correction introduces a difference between
the areal radii of the apparent horizon and the Schwarzschild radius as
\be\label{ah}
r(u, v_{ah}(u)) - a(u) \sim \mathcal{O}(\ell_p^2/a).
\ee
Here, $a(u)$ is the $u$-dependent Schwarzschild radius which decreases with time as \eqref{dota}, 
and we use $u_{ah}(v)$ and $v_{ah}(u)$ to denote the $u$ and $v$ coordinates
of the apparent horizon for given $v$ and $u$, respectively.

Our assumption on the energy-momentum tensor \eqref{pressure-bound}
is weaker than (certainly compatible with) what is assumed in the literature
for black holes with apparent horizons,
for which the energy-momentum tensor would be further constrained
by the regularity condition at the apparent horizon \cite{Baccetti:2018otf,Baccetti:2018qrp}.
Here, we aim at the most general energy-momentum tensor 
for the conclusion of the paper.

\section{Solving Semi-Classical Einstein Equations}

The semi-classical Einstein equation was solved in the near-horizon region
in Ref.\cite{HoMatsuoYokokura} with the assumption that
the vacuum energy-momentum tensor is given by those in Ref.\cite{Davies:1976ei}.
Here we generalize the result to any vacuum energy-momentum tensor satisfying
eq.\eqref{pressure-bound}.

\subsection{Near-Horizon Region}
\label{NHR}

We use the phrase ``near-horizon region'' to refer to
the region outside the collapsing matter
(with the areal radius $R_s(u)$)
but close to or inside the apparent horizon.
See Fig.\ref{Penrose}. 
In view of eqs.\eqref{r-a} and \eqref{pressure-bound},
we choose the outer boundary of the near-horizon region to be
\be
r(u, v) - a(u) = \frac{N\ell_p^2}{a(u)}
\label{boundary}
\ee
for an arbitrary large number $N$
satisfying $a^2/\ell_p^2 \gg N \gg 1$.
It follows eqs.\eqref{Sch-sol} and \eqref{boundary}
that the conformal factor in the metric \eqref{metric} 
around the outer boundary of the near-horizon region is
\be
C(u, v) \sim \mathcal{O}(\ell_p^2/a^2).
\label{C-order}
\ee
As we will see in the next subsection,
this condition is valid anywhere in the near-horizon region.

\hide{
\begin{figure}[h]
\vskip-4em
\center
\includegraphics[scale=0.5,bb=0 0 400 450]{AppHor11.pdf}
\vskip-1em
\caption{\small 
\new{
Penrose diagram
for the early stage of a dynamical black hole
(cropped at $u = u_1$).
Note that this neighborhood is included in both Fig.\ref{Penrose}(a) and Fig.\ref{Penrose}(b).
The ``near-horizon region'' defined in Sec.\ref{NHR} is the (blue shaded) region
bounded by the following 3 curves:
(1) the outer surface of the collapsing matter (thick green strip) from A to B,
(2) the curve of $u=u_1$ (dashed green line) from B to C,
(3) a curve outside but close to the time-like apparent horizon in vacuum from C to A.
We will study the proper length $\Delta L$ of a monotonic radial curve
connecting an arbitrary point on the surface of the collapsing matter between A and B
and another arbitrary point on the apparent horizon between A and C.
An example of such a curve (double arrowhead) is shown.
}
}
\label{AppHor-figure}
\vskip0em
\end{figure}
}

\begin{figure}[h]
\vskip-3em
\center
\includegraphics[scale=0.45,bb=0 0 500 400]{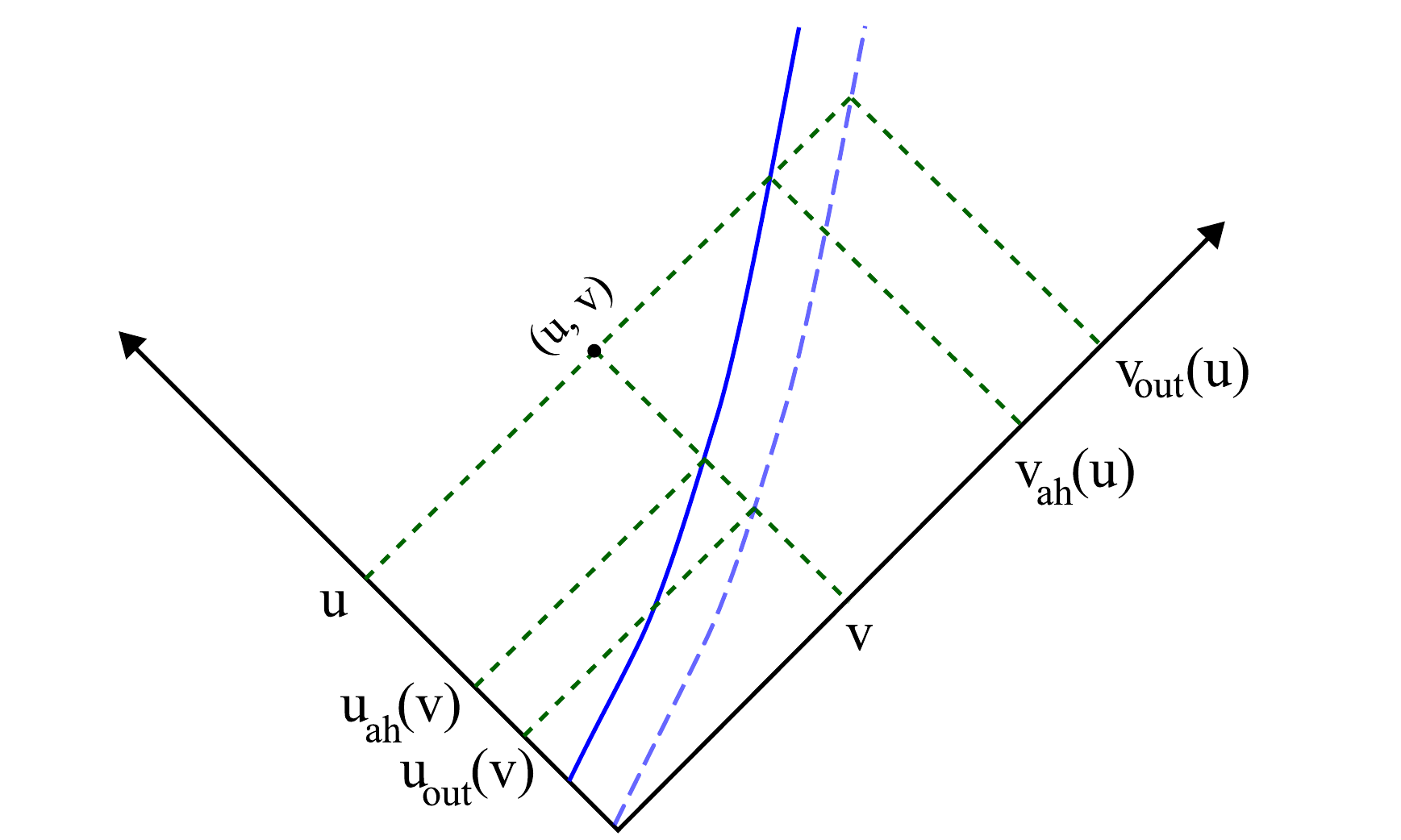}
\vskip1em
\caption{\small
The apparent horizon (solid blue) and the outer boundary of the near-horizon region (dash blue)
have their $u$, $v$ coordinates given by $u_{ah}(v)$, $v_{ah}(u)$
and by $u_{out}(v)$, $v_{out}(u)$, respectively.
The coordinates $(u, v)$ of a point inside the apparent horizon satisfy eq.\eqref{uvstar}.
}
\label{near-horizon-figure}
\vskip1em
\end{figure}

The trajectory of the outer boundary \eqref{boundary}
of the near-horizon region (see Fig.\ref{near-horizon-figure})
can be parameterized either by $u$ or by $v$.
Using $u$ as the parameter,
the $v$-coordinate on the trajectory
is denoted by $v_{out}(u)$.
Eq.\eqref{boundary} is satisfied with $v = v_{out}(u)$.
Conversely,
using $v$ as the parameter,
its $u$-coordinate is denoted by $u_{out}(v)$.
Clearly, $u_{out}$ and $v_{out}$ are the inverse functions of each other;
$u_{out}(v_{out}(u)) = u$ and $v_{out}(u_{out}(v)) = v$.

We focus our attention on a range of $u \in (u_0, u_1)$,
where $u_0$ is the moment when the apparent horizon emerges.
We emphasize that in the near-horizon region,
the ranges of $u$ and $v$ are both $\sim \mathcal{O}(a^3/\ell_p^2)$,
covering a huge space in terms of the $(u, v)$ coordinates.

The Schwarzschild radius $a(u)$ is monotonically decreasing due to Hawking radiation.
The maximal and minimal values of $a(u)$
from $u = u_0$ to $u_1$ are given by
\be
a_{max} = a(u_0)
\qquad \mbox{and} \qquad
a_{min} = a(u_1).
\ee
The ratio of the initial and final mass is denoted by
\be
n \equiv \frac{a_{max}}{a_{min}}.
\label{n-def}
\ee

Note that, for a point $(u, v)$ inside the apparent horizon, 
where the position of the apparent horizon for a given value of $u$ or $v$ 
is specified as $(u,v_{ah}(u))$ or $(u_{ah}(v),v)$ (see \eqref{ah}), respectively,
we have 
\be
v < v_{ah}(u) < v_{out}(u),
\qquad
u > u_{ah}(v) > u_{out}(v).
\label{uvstar}
\ee
\remove{since the apparent horizon is timelike 
and lies inside the outer boundary of the near-horizon region.}
\explanation{already said below}
(See Fig.\ref{near-horizon-figure}.)
These inequalities hold because the apparent horizon is timelike,
as a result of the negative quantum vacuum energy flow.

\subsection{Solution Of $C(u, v)$}

We solve $C(u,v)$ here from the semi-classical Einstein equation \eqref{SCEE} 
in the near-horizon region, 
with the boundary condition that it agrees with the Schwarzschild approximation.

Consider a particular combination of the semi-classical Einstein equations
\be
G^{\mu}{}_{\mu} - 6 G^{\th}{}_{\th}
= \kappa \left(\langle T^{\mu}{}_{\mu} \rangle - 6 \langle T^{\th}{}_{\th} \rangle\right).
\ee
It is equivalent to
\be
\del_u\del_v \Sigma(u, v) = \frac{C(u, v)}{4r^2(u, v)}
+ \frac{\kappa C(u, v)}{8} \left(\langle T^{\mu}{}_{\mu} \rangle 
- 6 \langle T^{\th}{}_{\th} \rangle\right),
\label{Sigma-eq}
\ee
where $\Sigma(u, v)$ is defined via
\be\label{sig}
C(u, v) \equiv \frac{e^{\Sigma(u, v)}}{r(u, v)}.
\ee

\begin{figure}[h]
\vskip-4em
\center
\includegraphics[scale=0.5,bb=0 -80 500 400]{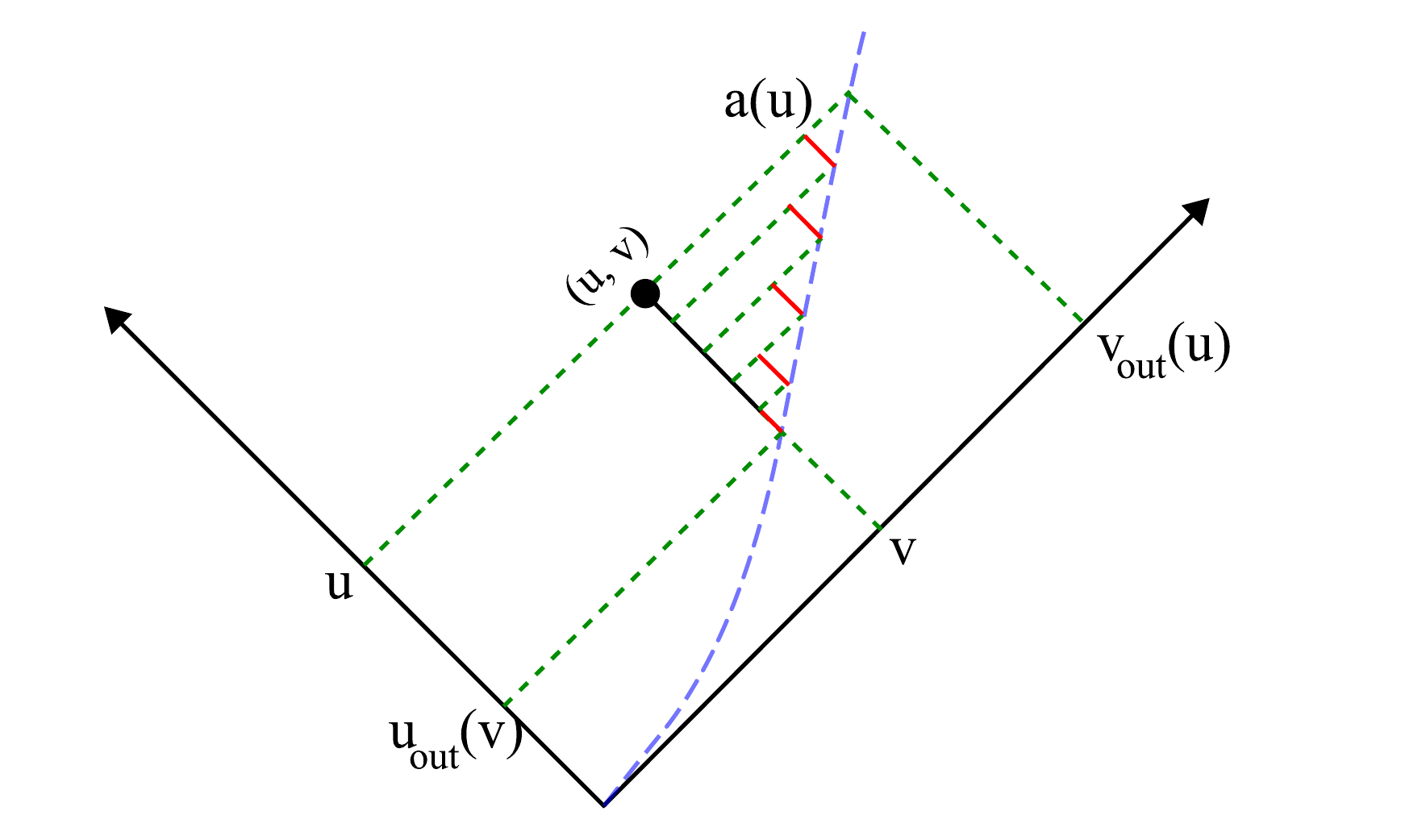}
\vskip-3em
\caption{\small
The dash blue curve represents the outer boundary of the near-horizon region.
Along the constant-$v$ curve,
the $u$-dependence of $\Sigma$ must agree with that of $\Sigma_0$
around the boundary \eqref{boundary}.
}
\label{Bu-figure}
\vskip1em
\end{figure}

For $r-a\sim \ell_p^2/a$, 
the naive order of magnitude of the left-hand side of eq.\eqref{Sigma-eq} is $\mathcal{O}(1/a^2)$,
and that of the right-hand side is $\mathcal{O}(\ell_p^2/a^4)$ 
because of eqs.\eqref{pressure-bound} and \eqref{C-order}.
Hence, the leading-order approximation of eq.\eqref{Sigma-eq} is 
\begin{equation}\label{Sigma-eq0}
\del_u\del_v \Sigma \simeq 0.
\end{equation}
It is solved by
\be
\Sigma(u, v) \simeq B(u) + \bar{B}(v)
\label{BBbar}
\ee
for arbitrary functions $B(u)$ and $\bar{B}(v)$.
To determine $B(u)$ and $\bar{B}(v)$,
we impose the boundary condition that
$C(u, v)$ matches with the Schwarzschild metric \eqref{Sch-sol}
around the outer boundary of the near-horizon region.
(See Fig.\ref{Bu-figure}.)

According to eq.\eqref{BBbar},
over an infinitesimal variation from $u$ to $(u + du)$ along a constant-$v$ curve,
the corresponding change in $\Sigma(u, v)$ is $du \partial_u B(u)$. 
Since the quantity $du \del_u\Sigma(u, v)$ is independent of $v$,
we take $v = v_{out}(u)$,
where the Schwarzschild solution is a good approximation
so that $\Sigma \simeq (v-u)/(2a) - 1 + \log(a)$
(see eq.\eqref{Sigma0} and Fig.\ref{Bu-figure}). 
We find
\be
du \partial_u B(u) \simeq du \del_u\Sigma(u, v)
\simeq du \del_u\Sigma_0(u, v') \Bigr|_{v'=v_{out}(u)} \simeq - \frac{du}{2a(u)}.
\ee
Here, we have used \eqref{Sigma0} with
\change{a time-dependent Schwarzschild radius} $a(u)$
and neglected contributions from $\partial_ua(u)$ as higher-order terms. 
Since the procedure above can be repeated for each infinitesimal segment $du$ for the same $v$, 
the equation above is immediately solved by
\be
B(u) \simeq B(u_{\ast}) - \int^u_{u_{\ast}} \frac{du'}{2a(u')}
\label{Bu}
\ee
for an arbitrary reference point $(u_{\ast}, v_{\ast})$ inside the near-horizon region. 
The function $a(u)$ can be interpreted as
the $u$-dependent Schwarzschild radius
for an infinitesimal slice from $u$ to $u+du$
at the outer boundary of the near-horizon region.
(See Fig.\ref{Bu-figure}.)

Similarly,
we have
\be
\bar{B}(v) \simeq \bar{B}(v_{\ast}) + \int^v_{v_{\ast}} \frac{dv'}{2\bar{a}(v')},
\label{Bv}
\ee
for some function $\bar{a}(v)$,
which is the $v$-dependent Schwarzschild radius
for an infinitesimal slice from $v$ to $v+dv$
at the outer boundary.
(For the static solution,
a single slice of the spacetime is enough to determine
the Schwarzschild radius.
Treating a small slice of the dynamical solution as a static one
would determine different Schwarzschild radii
for slices of fixed $u$ vs. fixed $v$.\footnote{
We can evaluate how different the two radii are 
by constructing the function $r(u,v)$ with a more specific condition of 
the energy-momentum tensor. 
See Appendix B of Ref.\cite{HoYokokura}.})

According to eqs.\eqref{Bu} and \eqref{Bv},
the 0-th order approximation of $\Sigma(u, v)$ is
\be
\Sigma(u, v) \simeq \Sigma(u_{\ast}, v_{\ast}) 
- \int_v^{v_{\ast}} \frac{dv'}{2\bar{a}(v')}
- \int^u_{u_{\ast}} \frac{du'}{2a(u')}. 
\label{Sigma-uv}
\ee
As a result, we obtain 
\begin{align}
C(u, v) &\simeq C(u_{\ast}, v_{\ast}) \frac{r(u_{\ast}, v_{\ast})}{r(u, v)}
\exp\left[- \int^u_{u_{\ast}} \frac{du'}{2a(u')} - \int_v^{v_{\ast}} \frac{dv'}{2\bar{a}(v')}\right].
\label{C-uv}
\end{align}
Going through the derivation of this solution,
one can check that
this formula can be applied to any two points $(u,v)$ and $(u_{\ast}, v_{\ast})$
in the near-horizon region. 

We shall adopt the convention of choosing
$(u_{\ast}, v_{\ast})$ to be on the outer boundary of the near-horizon region. 
To match with the Schwarzschild solution, 
we have from eq.\eqref{C-order}
\begin{align}
C(u_{\ast}, v_{\ast}) \sim \mathcal{O}(\ell_p^2/a^2(u_{\ast})).
\label{C0-bound}
\end{align}
In view of eqs.\eqref{uvstar} and \eqref{C-uv},
$C(u, v)$ becomes exponentially smaller
as we go deeper inside the near-horizon region.
As we will see below,
this is the crucial property of the near-horizon geometry
that prevents the distance between the collapsing matter
and the apparent horizon from becoming macroscopic
until the very late stage of the evaporation.

\subsection{First-Order Quantum Correction}
Since the large range of both $u$ and $v$ is $\mathcal O(a^3/\ell_p^2)$, 
the approximation \eqref{Sigma-eq0} might break down after integrating over $u$ or $v$. 
Here, we show that the approximation is valid over this large range of $u$ and $v$. 

The 1st-order quantum correction to $\Sigma(u, v)$ due to the right-hand side of eq.\eqref{Sigma-eq}
can be computed perturbatively as
\begin{align}
\Delta \Sigma(u, v) &\simeq
\int_{u_{\ast}}^u du' \int_{v_{\ast}}^v dv' \;
\left[\frac{1}{4r^2(u',v')}+\frac{\kappa}{8}(\langle T^{\mu}{}_{\mu} \rangle + 2\langle T^{\th}{}_{\th} \rangle)\right]
C(u', v')
\nn \\
&\lesssim
\left[\frac{1}{4r_{min}^2}+\frac{K}{8a_{min}^2}\right]C(u_{\ast}, v_{\ast})
\frac{r_{max}}{r_{min}}
\int_{u_{\ast}}^u du' \int_{v_{\ast}}^v dv' \;
e^{- \frac{v_{\ast}-v'}{2a_{max}}}
e^{- \frac{u'-u_{\ast}}{2a_{max}}}
\nn \\
&\lesssim
\left[\frac{a_{min}^2}{r_{min}^2}+\frac{K}{2}\right]\frac{a_{max}^2}{a_{min}^2}
\frac{r_{max}}{r_{min}} \, C(u_{\ast}, v_{\ast}),
\label{DeltaSigma}
\end{align}
where we have used eq.\eqref{uvstar} in the last step,
and $r_{max}$ and $r_{min}$ denote the maximal and minimal values of $r(u, v)$
in the near-horizon region, respectively.
We have also \remove{assumed}\change{used the condition} that
\be
|\langle T^{\mu}{}_{\mu} \rangle + 2\langle T^{\th}{}_{\th} \rangle| \leq \frac{K}{\kappa a^2}
\ee
holds any time
(here $K$ is of $\mathcal{O}(1)$),
according to our assumption \eqref{pressure-bound}.

The ratios $r_{max}/r_{min}$ and $a_{min}/r_{min}$ appearing in eq.\eqref{DeltaSigma}
can be estimated without an explicit functional form of $r(u,v)$. 
First, eqs.\eqref{ah} and \eqref{boundary} say that 
both $r(u, v_{ah}(u))$ and $r(u,v_{out}(u))$ approximately
equal the Schwarzschild radius $a(u)$ up to $\mathcal{O}(\ell_p^2/a)$ correction.
This is one of the two principles that govern the basic features of the function $r(u, v)$.
The other principle is simply 
the conditions $\partial_v r(u,v) < 0$ and $\partial_u r(u,v) < 0$, 
which hold by definition of the trapped region. 
Therefore, $r(u, v)$ decreases with $u$ for a fixed $v$. 
More precisely, $r(u, v)$ on each constant-$v$ curve decreases slower with $u$ for smaller values of $v$ 
because of the larger red-shift factor for deeper places inside the horizon. 
Thus, in the near-horizon region
the areal radius becomes the maximum on each constant-$v$ curve
at the outer boundary.
At the same time,
$r(u,v)$ decreases with $v$ for a fixed $u$. 
The minimum of the areal radius on each constant-$u$ curve
is that on the apparent horizon.

\begin{figure}[h]
\vskip-4em
\center
\includegraphics[scale=0.42,bb=0 -50 500 420]{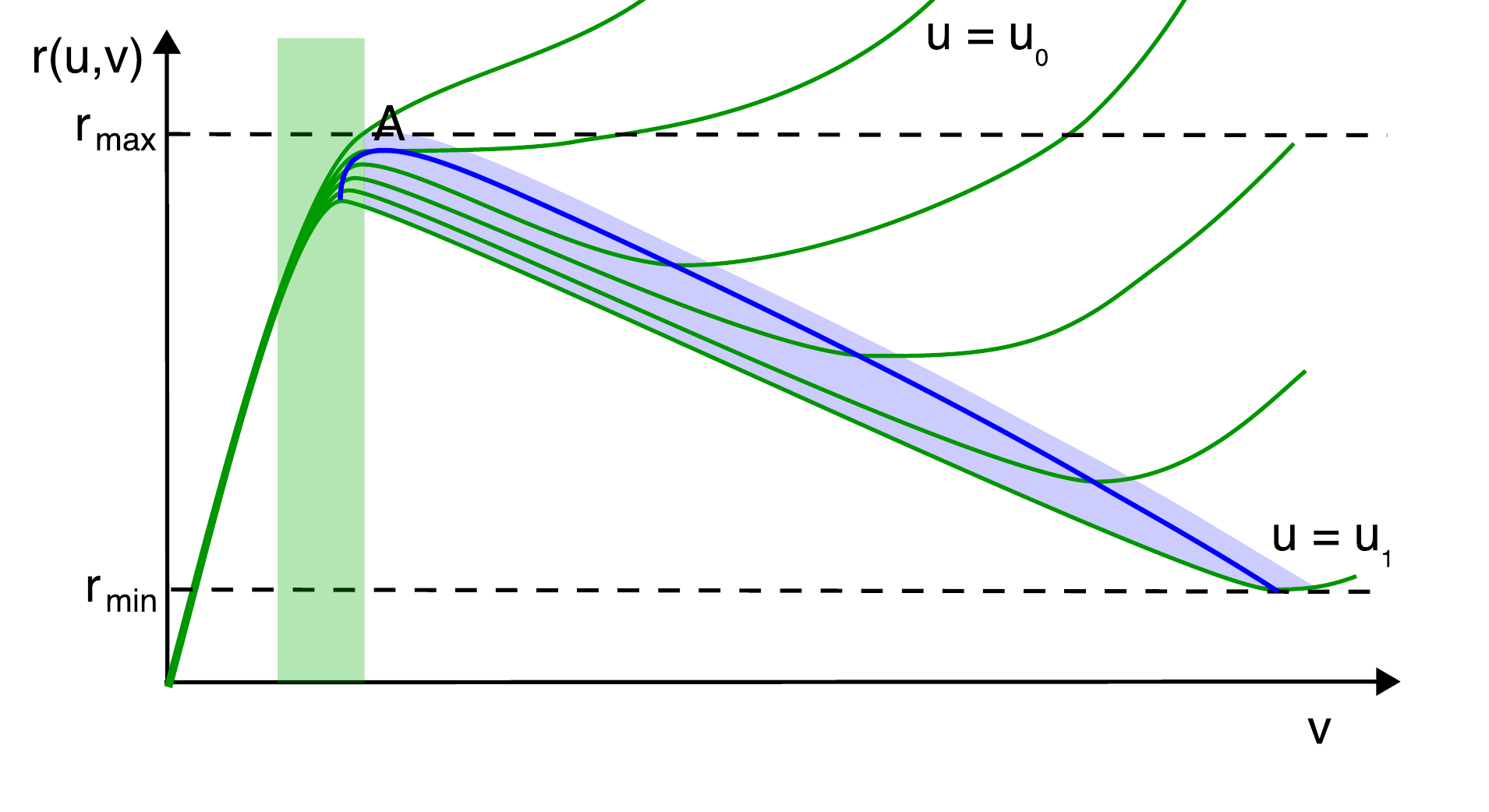}
\vskip-3em
\caption{\small
Schematic diagrams of $r(u, v)$ vs $v$.
For discrete values of $u$, including $u_0$ and $u_1$,
$r(u, v)$ is plotted as functions of $v$ (green curves) for given values of $u$.
These constant-$u$ curves almost coincide inside
the collapsing matter (thick green strip),
as well as in the flat spacetime inside the matter shell
(on the left of the green strip).
Outside the collapsing matter,
the local minima of the $r(u, v)$ curves at different $u$ (where $\partial_vr(u,v)=0$) 
are located on the apparent horizon $r(u_{ah}(v),v)$ (blue curve).
The near-horizon region is shown as the blue-shaded area.
}
\label{r-a-figure}
\vskip1em
\end{figure}

With the discussion above,
we can draw Fig.\ref{r-a-figure}, 
as a point-by-point image of Fig.\ref{Penrose}(c)
via the coordinate transformation from $(u, v)$ to $(r, v)$.
It should be clear that we can approximate $r_{max}$ and $r_{min}$, respectively, as 
the maximal and minimum Schwarzschild radii $a_{max}=a(u_0)$ and $a_{min}=a(u_1)$. 
Hence we have the relations
\be
\frac{r_{max}}{r_{min}} \simeq \frac{a_{max}}{a_{min}},
\qquad 
\frac{a_{min}}{r_{min}} \simeq 1.
\label{ratios-r-a}
\ee

We thus conclude from eqs.\eqref{DeltaSigma}, \eqref{C0-bound} and \eqref{ratios-r-a} that
\be\label{Dsigma}
\Delta\Sigma \lesssim \mathcal{O}\left(\frac{n^3\ell_p^2}{a_{\ast}^2}\right),
\ee
where $n$ is defined in eq.\eqref{n-def}.

In the definition of $\Sigma(u,v)$ \eqref{sig}, 
when we write the leading solution \eqref{C-uv} as $C^{(0)}$, 
the correction $\Delta \Sigma$ appears as 
$C=C^{(0)}e^{\Delta \Sigma}\approx C^{(0)}(1+\Delta \Sigma)$. 
Therefore,
the condition for $C$ to be dominated by $C^{(0)}$ is
\begin{equation}
\frac{C-C^{(0)}}{C^{(0)}}\approx \Delta \Sigma \ll 1.
\end{equation}
Thus eq.\eqref{Dsigma} means that whenever 
\be
n \ll \frac{a^{2/3}}{\ell_p^{2/3}}
\label{n-cond}
\ee
holds,  the 0-th order result \eqref{C-uv} is good.

\section{Distance in Near-Horizon Region}

\begin{figure}[h]
\vskip-2em
\center
\includegraphics[scale=0.6,bb=0 0 500 420]{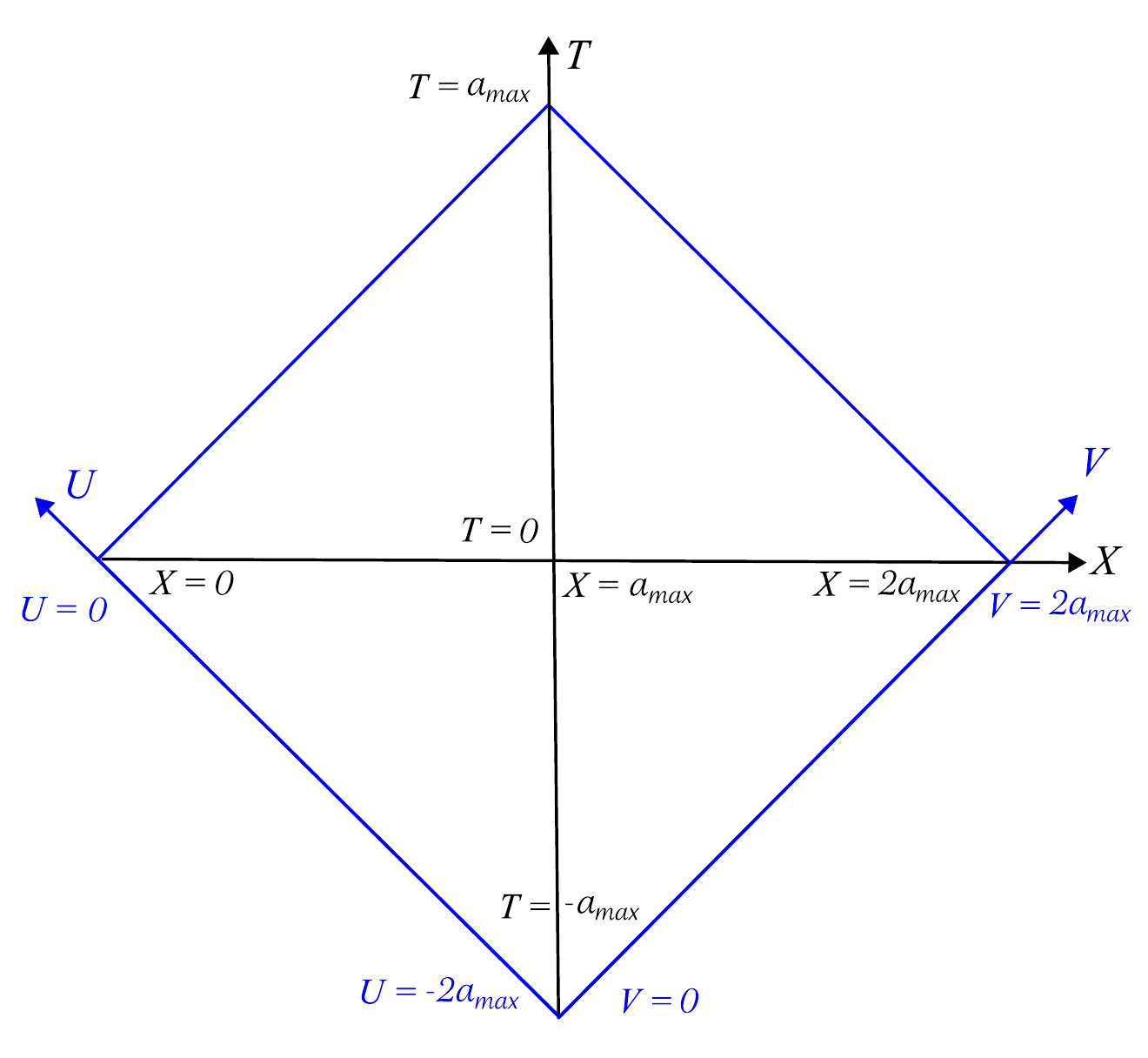}
\vskip0em
\caption{\small
The near-horizon region (the blue-shaded region in Fig.\ref{Penrose})
is mapped to a subspace of the $U-V$ plane
via eqs.\eqref{U-u} and \eqref{V-v}.
The distance between any two points in the near-horizon region is shorter
than the distance between their images in this diamond.
Corresponding to $U \in (-2a_{max}, 0)$ and $V \in (0, 2a_{max})$,
we have $X \in (0, 2a_{max})$ and $T \in (-a_{max}, a_{max})$.
Within this space,
the curve with maximal proper distance is the $X$-axis at $T = 0$,
and the curve with the maximal proper time is the $T$-axis at $X = a_{max}$.
}
\label{UV-space}
\vskip1em
\end{figure}

Define new (Kruskal-like) coordinates (see Fig.\ref{UV-space}) as
\begin{align}
U &\equiv T-X = - 2a_{max}e^{-\frac{u-u_{\ast}}{2a_{max}}} \; \in \; (-2a_{max}, 0),
\label{U-u}
\\
V &\equiv T+X = 2a_{max}e^{-\frac{v_{\ast}-v}{2a_{max}}} \; \in \; (0, 2a_{max}),
\label{V-v}
\end{align}
where the range is deduced from the inequalities \eqref{uvstar}. 
The proper length along a spacelike curve ${\cal C}$ restricted to the $u-v$ plane
\footnote{
Each point in the $u-v$ plane represents a 2-sphere in 4D.
The curve ${\cal C}$ is restricted to the radial and temporal directions
to define the distance between two concentric 2-spheres.
}
is
\begin{align}
\Delta L &= \int_{\cal C} \sqrt{- C(u, v) du dv}
\leq C^{1/2}(u_{\ast}, v_{\ast})\frac{r^{1/2}(u_{\ast},v_{\ast})}{r^{1/2}(u,v)} \int_{\cal C}
\sqrt{- e^{-\frac{u-u_{\ast}}{2a_{max}}} e^{-\frac{v_{\ast}-v}{2a_{max}}} dudv}
\nn \\
&
\leq C^{1/2}(u_{\ast}, v_{\ast})\frac{r^{1/2}_{max}}{r^{1/2}_{min}} \int_{\cal C} \sqrt{- dUdV}
\simeq C^{1/2}(u_{\ast}, v_{\ast})\frac{a^{1/2}_{max}}{a^{1/2}_{min}} \int_{\cal C} \sqrt{dX^2 - dT^2}
\nn \\
&\leq C^{1/2}(u_{\ast}, v_{\ast})\frac{a^{1/2}_{max}}{a^{1/2}_{min}} \int_{\cal C}dX
\leq 2\frac{a^{3/2}_{max}}{a^{1/2}_{min}}C^{1/2}(u_{\ast}, v_{\ast}) 
\sim
\mathcal{O}(n^{3/2}\ell_p),
\label{bound-DL}
\end{align}
where we have used eqs. \eqref{C-uv}, \eqref{C0-bound},
and \eqref{ratios-r-a} and the definition $n \equiv a_{max}/a_{min}$.

Via a similar calculation,
the proper time for a timelike curve is also
at most of the same order of magnitude.

This result means that any proper distance in the radial direction
inside the near-horizon region is bounded from above by $n^{3/2}\ell_p$. 
In particular,
this conclusion applies to the proper distance
between the apparent horizon and the surface of the collapsing matter, 
since both are in the near-horizon region. 

At first glance, this result might seem purely classical 
because the 0-th order solution $C(u,v)$ \eqref{C-uv} has been obtained from eq.\eqref{Sigma-eq0},
which includes no quantum correction. 
However, we have arrived at eq.\eqref{bound-DL} by using the inequalities \eqref{uvstar},
which is a consequence of the negative vacuum energy. 
Hence this result actually relies on the quantum effect.

At the Page time ($n = 2$),
the distance between the collapsing matter and the apparent horizon is at most $\ell_p$.
When $99\%$ of the black hole is evaporated,
the distance is at most $100$ times the Planck length. 
\footnote{Interestingly,
another self-consistent model \cite{Kawai:2013mda,KMY}, 
in which there is no trapped region, 
provides a similar result. 
A collapsing matter is just \textit{above} the Schwarzschild radius by a Planckian distance. 
In this sense, the conventional model studied in the present paper
might eventually be closely related to such a model.}

A peculiar feature of our estimate of the upper bound $\Delta L$
for the proper distance in the near-horizon region is that
$\Delta L$ does not depend on both the initial and final masses,
but only on their ratio.
\footnote{
It is also independent of other details such as
when the near-horizon region appears
and how long it takes the black hole to evaporate to $1/n$ of its initial mass.
}
As a result,
the upper bound $\Delta L$ is the same for black holes that
could be dramatically different in size,
as long as they have evaporated to the same portion of their initial mass.
This is actually a coincidence in the sense that,
if we repeat the calculation above for a higher spacetime dimension $D > 4$,
the estimate $\Delta L$ would no longer have this property.
Instead,
one has \cite{Ho:2020cvn}
\begin{align}
C \simeq C_{\ast} e^{- (D-3)(u - u_{\ast} + v_{\ast} - v)/2a_{\ast}},
\qquad
C_{\ast} \sim \mathcal{O}(\ell_p^{D-2}/a_{\ast}^{D-2}),
\end{align}
so that the upper bound on the distance is now
\begin{align}
\Delta L \sim a_{\ast}C_{\ast}^{1/2} \sim \ell_p^{(D-2)/2}/a_{\ast}^{(D-4)/2}.
\end{align}
Hence, 
$\Delta L$'s independence of $a_{\ast}$ (or $M_{\ast}$) is
only a coincidence for 4D black holes.

To summarize,
we have shown that,
after the collapsing matter enters the apparent horizon
and before the black hole's mass is reduced to $1/n$ of its initial mass,
the proper distance $\Delta L$ between the surface of the collapsing matter
and the timelike apparent horizon is bounded from above by 
$\Delta L \leq n^{3/2} \ell_p$. 
This bound holds for all paths along which $d\th = d\phi = 0$
as long as $n \ll a^{2/3}/\ell_p^{2/3}$.
This result is very interesting because it shows that,
from the viewpoint of a low-energy effective theory
below the Planckian scale,
it is strictly speaking indistinguishable whether
the matter has entered the apparent horizon in this regime.
Our result is in agreement with Refs.\cite{Baccetti:2018otf,Baccetti:2018qrp},
which investigated the implications of the violation 
of the null energy condition in the near-horizon geometry,
and questioned the validity of the low-energy effective theory
in the near-horizon regime.
Furthermore, 
we expect to derive from this result further implications 
to the information paradox \cite{HoYokokura}.

\section*{Acknowledgement}
We thank Hikaru Kawai for valuable discussions and suggestions.
P.M.H. thanks iTHEMS at RIKEN,
University of Tokyo and Kyoto University
for their hospitality during his visits where the major part of this work was done.
P.M.H.\ is supported in part by the Ministry of Science and Technology, R.O.C. 
and by National Taiwan University. 
The work of Y.M.\ is supported in part by
JSPS KAKENHI Grants No.~JP17H06462 and JP20K03930.
Y.Y.\ is partially supported by Japan Society of Promotion of Science (JSPS), 
Grants-in-Aid for Scientific Research (KAKENHI) Grants No.\ 18K13550 and 17H01148. 
Y.Y.\ is also partially supported by RIKEN iTHEMS Program.

\appendix

\section{Schwarzschild Approximation}
\label{review}

As long as $a \gg \ell_p$,
the Schwarzschild solution with a time-independent
Schwarzschild radius $a$
should be a good approximation
within a time scale $\lesssim \mathcal{O}(a)$
at a place well outside the Schwarzschild radius 
where
\footnote{
Strictly speaking,
$r - a \gtrsim N\ell_p^2/a$ is satisfied 
both outside and inside the apparent horizon.
We refer to the region outside the Schwarzschild radius here.
}
\be
r - a \gtrsim \frac{N\ell_p^2}{a}
\label{well-outside}
\ee
with a sufficiently large (but finite) $N$
(e.g. $N \sim 10000$).

We consider the small neighbourhood outside the apparent horizon
\footnote{
The choice of the domain $\left(\frac{N\ell_p^2}{2a}, \frac{N\ell_p^2}{a}\right)$ is arbitrary, 
as long as it covers a neighbourhood of eq.\eqref{well-outside}
where the Schwarzschild approximation is good.
}
\be
r - a \in \left(\frac{N\ell_p^2}{2a}, \frac{N\ell_p^2}{a}\right)
\label{boundary-strip}
\ee
and a period of time $[t_0,t_0+\Delta t]$ where $\Delta t\sim \mathcal{O}(a)$. 
We demand that $N$ is sufficiently large
so that the Schwarzschild approximation is good,
but not too large ($\frac{a^2}{\ell_p^2} \gg N$)
so that $(r - a)/a \ll 1$.
The metric in the region \eqref{boundary-strip}
can be approximated as the usual Schwarzschild metric 
with a constant radius $a = a(t_0)$: 
\be\label{Schmetric}
ds^2 = - \left(1 - \frac{a}{r(u, v)}\right) du dv + r^2(u, v) d\Omega^2,
\ee
where the areal radius $r(u, v)$ is related to
the tortoise coordinate $r^{\ast}$ via
\begin{align}
r^{\ast} \equiv \frac{v - u}{2}
= r(u, v) + a\log\left(\frac{r(u, v)}{a} - 1\right)
\simeq a + a \log\left(\frac{r(u, v) - a}{a}\right).
\label{rstar-def}
\end{align}
In the neighbourhood \eqref{boundary-strip}, 
the Schwarzschild metric \eqref{Schmetric} becomes approximately 
\be
ds^2 \simeq - C_0(u, v) du dv + a^2 d\Omega^2,
\qquad
\mbox{where}
\qquad
C_0(u, v) \equiv \frac{a}{r} \, e^{\frac{v-u-2a}{2a}}.
\label{C0}
\ee
In the neighbourhood \eqref{boundary-strip}, 
the metric \eqref{Schmetric} means 
$C(u, v) \sim \mathcal{O}(\ell_p^2/a^2)$.
Then, $\Sigma_0$,
which is defined by eq.\eqref{sig}, is 
\be
\Sigma_0(u, v) \simeq \frac{v-u}{2a} - 1 + \log(a).
\label{Sigma0}
\ee


\end{document}